\documentclass[]{iopart}
\usepackage[latin1]{inputenc}
\usepackage{graphicx}

\newcommand{\avg}[1]{E\,#1}

\begin{document}

\title{Bond chaos in the Sherrington-Kirkpatrick model}
\author{T Aspelmeier}
\address{Max Planck Institute for Dynamics and Self Organization,
37073 Göttingen, Germany}

\begin{abstract}
We calculate the probability distribution of the overlap between a spin glass
and a copy of itself in which the bonds are randomly perturbed in varying
degrees. The overlap distribution is shown to go to a $\delta$
distribution in the thermodynamic limit for arbitrarily small perturbations
(bond chaos) and we obtain the scaling behaviour of
the distribution with system size $N$ in the high and low temperature phases 
and exactly at the critical temperature. The results are relevant
for the free energy fluctuations in the Sherrington-Kirkpatrick model
\cite{Aspelmeier:2007a}.
\end{abstract}


%\submitto{J. Phys. A}

%\maketitle

\section{Introduction}

Chaos is one of the fascinating properties of spin glasses. Whenever a spin glass
is subjected to a change (e.g.\ a change in temperature or in the coupling
constants), the equilibrium state of the system changes completely (in the
thermodynamic limit), no matter how small the change. This behaviour has been
first observed in hierarchical models \cite{McKay:1982} and was then suggested
for finite dimensional spin glasses \cite{Bray:1987} within the droplet theory.
Chaos is however not restricted to replica symmetric droplet-like scenarios: in
\cite{Rizzo:2003} it was shown that temperature chaos exists in the
Sherrington-Kirkpatrick model \cite{Sherrington:1975} which breaks the replica
symmetry. In this paper we show that bond chaos also exists in the
Sherrington-Kirkpatrick model. This is not surprising since bond chaos is usually
a stronger effect than temperature chaos \cite{Ney-Nifle:1998}. However, it has
recently been found that an intricate connection exists between bond chaos and
the sample-to-sample fluctuations of the free energy
\cite{Aspelmeier:2007a,Aspelmeier:2008cpre}, which is a long standing problem in
spin glass and extreme value theory. It is therefore necessary to calculate bond
chaos in order to make progress on fluctuations, and in this paper we will
present the details of the calculation which was only sketched in
\cite{Aspelmeier:2007a}.

Temperature chaos has been simulated extensively in the literature, see e.g.\
\cite{Katzgraber:2007} and references therein. Bond chaos has not been simulated
quite so much but nevertheless for many different models, including the
Edwards-Anderson model in dimensions $1$ to $4$
\cite{Ney-Nifle:1998,Billoire:2000,Krzakala:2005,Sasaki:2005,Katzgraber:2007}, on
the hierarchical Berker lattice
\cite{Banavar:1987,Aspelmeier:2002b} and on Bethe lattices \cite{Krzakala:2005}
but, to the best of our knowledge, never for the Sherrington-Kirkpatrick model.
Here we will show that the probability distribution of the overlap, which is our
main quantity of interest, has a relatively complicated finite size scaling
behaviour (see Eq.~(\ref{probbelow}) below).

This paper is organized as follows. In Sec.~\ref{secmodel} we specify the model
and method we will be using and explain in Sec.~\ref{secprobability} how to
obtain the probability distribution of the overlap for this model. We then
proceed with the replica calculation in Sec.~\ref{secreplica} where we explicitly
calculate this distribution above, at and below the critical temperature.
% These predictions are compared to computer simulations in
% Sec.~\ref{secnumerical}, and we
We end with a conclusion in Sec.~\ref{secconclusion}.

\section{Model}
\label{secmodel}

We will compare two real replicas of a Sherrington-Kirkpatrick spin glass
with Hamiltonian
\begin{equation}
\mathcal H_\epsilon = -\frac{1}{\sqrt{N}}\sum_{i<j} K_{ij}(\epsilon) s_i s_j,
\end{equation}
where the $s_i$ ($i=1,\dots,N$) are $N$ Ising spins and
\begin{equation}
K_{ij}(\epsilon) = \frac{1}{\sqrt{1+\epsilon^2}}J_{ij} +
\frac{\epsilon}{\sqrt{1+\epsilon^2}}J'_{ij}
\end{equation}
are Gaussian random variables of unit variance, composed of
independent Gaussian random variables $J_{ij}$ and $J'_{ij}$, also of unit variance.
The parameter $\epsilon$ is a measure of ``distance'' between the sets of bonds
$\{K_{ij}(0)\}$ and $\{K_{ij}(\epsilon)\}$. If $\epsilon=0$, the bonds are
equal, if $\epsilon=\infty$, the bonds are completely uncorrelated.

Our main question concerns the disorder averaged probability distribution of the
spin-spin overlap $P_\epsilon(q)$ between the two replicas, the first with
$\mathcal H_0$ and the other with $\mathcal H_\epsilon$. Here $q$ is defined by
\begin{equation}
q = \frac 1N \sum_i s_i^{1,0} s_i^{2,\epsilon},
\end{equation}
where $s_i^{r,x}$ is the $i$th spin in replica $r$, which has Hamiltonian
$\mathcal H_x$.

We know that for temperature chaos (i.e.\ when comparing two replicas with
identical bonds but at different temperatures), even an infinitesimal temperature
difference $\Delta T$ leads to $P_{\Delta T}(q)=\delta(q)$ \textit{in the
thermodynamic limit} \cite{Rizzo:2003}. This means that equilibrium states in the
two replicas are totally uncorrelated. Since bond chaos is usually a stronger
effect than temperature chaos, we expect the same here, and we will show this
explicitly below. More interesting, however, and also more important, is the
question of how $P_\epsilon(q)$ scales towards the $\delta$-function with system
size $N$. This information is for instance needed for the calculation of the
sample-to-sample free energy fluctuations
\cite{Aspelmeier:2007a,Aspelmeier:2008cpre}.

\section{Probability distribution of the overlap}
\label{secprobability}

In order to calculate $P_\epsilon(q)$ we first try to formally calculate
$P_{\epsilon,J}(q)$, the nonaveraged probability distribution to find the overlap
$q$. Here and in the following, the subscript $J$ indicates a nonaveraged
quantity. Given a realization of the disorder for the two replicas, we can write
down a partition function $Z_{\epsilon,J}(q)$ for them, constrained to have the
overlap $q$,
\begin{equation}
\fl
Z_{\epsilon,J}(q) = \mathrm{Tr}\,\delta\left(q-\frac 1N \sum_i s_i^{1,0}
s_i^{2,\epsilon}\right)\exp\left( \beta \sum_{i<j}K_{ij}(0) s_i^{1,0} s_j^{1,0}
+ \beta \sum_{i<j}K_{ij}(\epsilon) s_i^{2,\epsilon} s_j^{2,\epsilon} \right).
\end{equation}
This method of constraining two systems to have a given overlap was
first suggested in \cite{Franz:1992}.
From this one gets the free energy $\beta F_{\epsilon,J}(q) = -\log
Z_{\epsilon,J}(q)$, and defining $Y_{\epsilon,J}=\int_0^1
dq\,Z_{\epsilon,J}(q)$, the probability distribution of $q$ is given by a Boltzmann factor,
\begin{equation}
P_{\epsilon,J}(q) = \frac{e^{-\beta F_{\epsilon,J}(q)}}{Y_{\epsilon,J}} =
\frac{Z_{\epsilon,J}(q)}{Y_{\epsilon,J}}.
\end{equation}
Of course, we cannot calculate $F_{\epsilon,J}(q)$ for a given disorder but we
can calculate its disorder average $F_\epsilon(q)=\avg{F_{\epsilon,J}}$ (the
disorder average is denoted by the symbol $\avg{\cdots}$)
in the thermodynamic limit by replica
methods. This will be done below. But first we want to continue formally in
order to assess the approximations we are going to make. 

We split $F_{\epsilon,J}(q)$ into three parts,
\begin{equation}
F_{\epsilon,J}(q) = N f_\epsilon(q) + \Delta F_\epsilon(q) + \Delta
F_{\epsilon,J}(q).
\end{equation}
The first part, $Nf_\epsilon(q)$, is the extensive part of the
average free energy. The second part, $\Delta F_\epsilon(q)$, is the finite
size correction to it. Finally $\Delta F_{\epsilon,J}(q)$ is the
fluctuation of the free energy about its disorder average $F_\epsilon(q)$, 
so $\avg{\Delta F_{\epsilon,J}(q)}=0$.

In order to calculate $P_\epsilon(q) = \avg{P_{\epsilon,J}(q)}$, we first look at
$Y_{\epsilon,J}$. It can be written as follows,
\begin{eqnarray}
Y_{\epsilon,J} &= \int_0^1 dq\,e^{-\beta N f_\epsilon(q)-\beta(\Delta
F_\epsilon(q)+\Delta F_{\epsilon,J}(q))} \nonumber\\
&= \int_0^1 dq\,e^{-\beta N f_\epsilon(q)}\left(
1+\sum_{n=1}^\infty \frac{(-\beta)^n}{n!}
\frac{\int_0^1 dq\,e^{-\beta N f_\epsilon(q)}(\Delta F_\epsilon(q)+\Delta
F_{\epsilon,J}(q))^n}{\int_0^1 dq\,e^{-\beta N f_\epsilon(q)}}
\right)\nonumber\\
&= Y_\epsilon^0 \left(
1+\sum_{n=1}^\infty \frac{(-\beta)^n}{n!} \left[(\Delta F_\epsilon+
\Delta F_{\epsilon,J})^n \right]_0 \right),
\end{eqnarray}
where $Y_\epsilon^0 := \int_0^1 dq\,e^{-\beta N f_\epsilon(q)}$ and
$[\cdots]_0$ is the average taken with the probability distribution
\begin{equation}
P_\epsilon^0(q) = \frac{e^{-\beta N f_\epsilon(q)}}{Y_\epsilon^0}.
\end{equation}
We then have
\begin{equation}
\frac{1}{Y_{\epsilon,J}} = \frac{1}{Y_\epsilon^0}\int_0^\infty dx\, 
e^{-x(1+\sum_{n=1}^\infty \frac{(-\beta)^n}{n!} \left[(\Delta F_\epsilon+
\Delta F_{\epsilon,J})^n \right]_0)}
\end{equation}
and
\begin{eqnarray}
\fl
P_\epsilon(q) = P_\epsilon^0(q) E \int_0^\infty dx\,
e^{-x -x \sum_{n=1}^\infty \frac{(-\beta)^n}{n!} \left[(\Delta F_\epsilon+
\Delta F_{\epsilon,J})^n \right]_0 -\beta (\Delta F_\epsilon(q)+\Delta
F_{\epsilon,J}(q))} \\
= P_\epsilon^0(q) \Big(
1-\beta(\Delta F_\epsilon(q)-[\Delta F_\epsilon]_0) \nonumber\\
-
\frac{\beta^2}{2} \left[(\Delta F_\epsilon(q)-[\Delta F_\epsilon]_0)^2\right]_0 -
\frac{\beta^2}{2} E\, \left[(\Delta F_{\epsilon,J}(q)-[\Delta
F_{\epsilon,J}]_0)^2\right]_0 \nonumber\\
 + \frac{\beta^2}{2}\left(\Delta F_\epsilon(q)-[\Delta
F_\epsilon]_0\right)^2 + \frac{\beta^2}{2}E\,\left(\Delta
F_{\epsilon,J}(q)-[\Delta F_{\epsilon,J}]_0\right)^2 + \cdots \Big).
\label{probability}
\end{eqnarray}
The last line follows from expanding the exponentials and sorting by powers of
$\Delta F_\epsilon$ and $\Delta F_{\epsilon,J}$.

Eq.~(\ref{probability}) shows precisely what can be expected of the calculation
which is to follow. We will calculate the first term of this expression,
$P_\epsilon^0(q)$. This is the same term one would have written down from the
start by following large deviation statistics principles. Anything beyond this
would require us to calculate the finite size corrections $\Delta F_\epsilon(q)$
and $\Delta F_{\epsilon,J}(q)$ which is impossible due to the massless modes
present in the spin glass phase. But our precise analysis allows us to gauge the
applicability of the large deviation statistics approximation: although $\Delta
F_\epsilon(q)$ and $\Delta F_{\epsilon,J}(q)$ are not necessarily small (they
grow with some subdominant power of $N$), they only appear in the combinations
$\Delta F_\epsilon(q)-[\Delta F_\epsilon]_0$ and $\Delta
F_{\epsilon,J}(q)-[\Delta F_{\epsilon,J}]_0$. If these differences are small, our
results not only hold for ``large'' deviations but also for ``small'' ones. While
we have no proof for this, application of our results to the free energy
fluctuations above and at the critical temperature
\cite{Aspelmeier:2007a,Aspelmeier:2008cpre} shows that it is at least true at
those temperatures. As to the low temperature phase, we will see later that for
$\epsilon=0$ and $q$ less than the Edwards Anderson order parameter
$q_{\mathrm{EA}}$, the corrections are of order $1$ but small enough not to
introduce any qualitative changes, and we expect the same to hold for
$\epsilon>0$.

\section{Replica calculation}
\label{secreplica}

Temperature chaos in mean-field spin glasses has been treated in the literature
\cite{Rizzo:2001,Rizzo:2003} and we refer the reader to these papers for details.
Repeating Rizzo's calculation \cite{Rizzo:2001} (see also \cite{Billoire:2003}
for the calculation at $\epsilon=0$) but for bond chaos rather than temperature
chaos, one arrives at the following expression for the disorder averaged,
replicated, constrained partition function $E\,Z_{\epsilon,J}^n(q)$
\begin{eqnarray}
\fl
E\,Z_{\epsilon,J}^n(q) = \int\left(\prod_{\alpha=1}^n
dz_\alpha\right) \left(\prod_{\alpha<\beta}^{2n}dT_{\alpha\beta}\right)
\exp\left(2N\left[\frac\tau 2 \sum_{\alpha\beta}^n Q_{\alpha\beta}^2 + 
\frac{\tau'}{2}\sum_{\alpha\beta}^n R_{\alpha\beta}^2\right.\right.\nonumber\\
+\frac{w}{6}\left(\sum_{\alpha\beta\gamma}^n Q_{\alpha\beta}Q_{\beta\gamma}Q_{\gamma\alpha} + 
3\sum_{\alpha\beta\gamma}^n Q_{\alpha\beta}R_{\beta\gamma}R_{\gamma\alpha} \right)
+\frac{y}{12}\left(\sum_{\alpha\beta}^n Q_{\alpha\beta}^4 +
\sum_{\alpha\beta}^n R_{\alpha\beta}^4\right)\nonumber\\
 -q\sum_{\alpha=1}^n
z_\alpha\left.\left.+\frac{\sqrt{1+\epsilon^2}}{4}\sum_{\alpha=1}^n
\left(R_{\alpha\alpha}^2-(R_{\alpha\alpha}-2z_\alpha)^2\right)\right]\right).
\label{replicahamiltonian}
\end{eqnarray}
Here, $\tau=\frac 12(1-1/\beta^2)$, $\tau'=\frac 12
(1-\sqrt{1+\epsilon^2}/\beta^2)$ and the $2n\times 2n$ matrix $T$ is given by
\begin{equation}
T=\left(\begin{array}{cc}
Q & R \\ R & Q
\end{array}
\right)
\end{equation}
and the $n\times n$ matrix $Q$ is zero on the diagonal, while $R$, also of size
$n\times n$, is not. The latter will therefore be split into a part on the
diagonal $p_d\mathbf{1}$ and the rest $P$. At this stage it is an ansatz to
make all entries on the diagonal of $R$ equal to $p_d$ but this need not be done
and one could proceed without this assumption and justify it later.

Note that, in contrast to \cite{Rizzo:2001}, we are using the truncated model
here by keeping only those fourth order terms which are written down in
Eq.~(\ref{replicahamiltonian}). This will make the calculations below
much easier.

The saddle point equations following from this expression can easily be derived
and, making a Parisi symmetry breaking ansatz for $P$ and $Q$, one gets the
following system of equations:
\begin{eqnarray}
\fl
0 = \tau q(x) + \frac y3 q^3(x) - \frac w2 \int_0^x dz\,(q(x)-q(z))^2 - 
\frac w2 \int_0^x dz\,(p(x)-p(z))^2 \nonumber\\
- w(\overline{p}-p_d)p(x) -
w\overline{q}q(x)
	\label{saddle1}
\\
\fl
0 = \tau' p(x)+ \frac y3 p^3(x) -w\int_0^x dz\,(p(x)-p(z))(q(x)-q(z))\nonumber\\
- w(\overline{p}-p_d)q(x) - w\overline{q}p(x)
\label{saddle2}
\\ 
\fl
q = p_d + \frac{2y}{3}p_d^3 - 2w\int_0^1 dz\,p(z)q(z).
\label{saddle3}
\end{eqnarray}
The variable $q$ (without argument) in the last equation is the overlap the two
real replicas are forced to have. As usual, $\overline p$ and $\overline q$
denote the integrals over $p(x)$ and $q(x)$ from 0 to 1. The free energy per
spin may be expressed in terms of the solutions $p(x)$, $q(x)$ and $p_d$ of
these equations and is given by
\begin{eqnarray}
\fl
\beta f_\epsilon(q) = qp_d-\frac{p_d^2}{2} - \frac{yp_d^4}{6} -
\frac{q^2}{2(1-2\tau')} + \tau\int_0^1 dz\,q^2(z) + \frac y6 \int_0^1 dz\,q^4(z)\nonumber\\
 -\frac w3\int_0^1 dz\,zq^3(z) - w\int_0^1 dz\,q^2(z)\int_z^1 dz'\,q(z')+2wp_d\int_0^1 dz\,p(z)q(z)\nonumber\\
 + \tau'\int_0^1 dz\,p^2(z) + \frac y6 \int_0^1 dz\,p^4(z) - w\int_0^1 dz\,p^2(z)\int_z^1 dz'q(z')\nonumber\\
-2w\int_0^1 dz\,p(z)q(z)\int_z^1 dz'\,p(z')-w\int_0^1 dz\,zp^2(z)q(z).
\label{free}
\end{eqnarray}
It can be checked that this free energy gives back the saddle point equations
Eqs.~(\ref{saddle1})--(\ref{saddle3}) when the derivatives with respect to
$p(x)$, $q(x)$ and $p_d$ are taken.

\subsection{Above the critical temperature}

Above the critical temperature, $\tau<0$, the saddle point equations can be
solved perturbatively in the limit of small $q$. For $q=0$, the exact solution
is $q(x)=p(x)=0$ and $p_d=0$. For $0<q\ll |\tau|$ it is easy to see that
$q(x)=\mathcal O(q^2)$, $p(x)=\mathcal O(q^2)$ and $p_d=q+\mathcal O(q^3)$.
Plugging this into the free energy, Eq.~(\ref{free}), yields
\begin{equation}
\beta f_\epsilon(q) = \frac{q^2}{2}\left(
1-\frac{\beta^2}{\sqrt{1+\epsilon^2}}\right) + \mathcal O(q^4).
\label{freeabove}
\end{equation}
The probability distribution $P_\epsilon^0(q)$ is thus 
\begin{equation}
P_\epsilon^0(q) \propto e^{-N(\frac{q^2}{2}h(\epsilon)+\mathcal O(q^4))}
\end{equation}
with $h(\epsilon)=1-\frac{\beta^2}{\sqrt{1+\epsilon^2}}$.

\subsection{At the critical temperature}

At the critical temperature, $\tau=0$, we can solve the saddle point equations
in the limit $q\ll |\tau'|$. Just like above the critical temperature, the
solution is $q(x)=\mathcal O(q^2)$, $p(x)=\mathcal O(q^2)$ and $p_d=q+\mathcal
O(q^3)$, and the free energy is also given by Eq.~(\ref{freeabove}) in this
limit. Expanding in powers of $\epsilon$, we find
\begin{equation}
\beta f_\epsilon(q) =
\frac{q^2}{2} h(\epsilon) =
\frac{q^2\epsilon^2}{4}+\mathcal O(\epsilon^4) \qquad(q\ll\epsilon^2\ll 1).
\label{freeat1}
\end{equation}
We will also need the solution of the saddle point equations in a different
limit, namely $\epsilon^2\ll q\ll 1$. To obtain this, it is convenient to
rewrite the saddle point equations and the free energy in terms of the
functions $a(x):=q(x)+p(x)$ and $b(x):=q(x)-p(x)$. A straightforward
calculation leads to 
\begin{eqnarray}
\fl
0 = (\tau+w p_d) a(x) - \frac{\tau-\tau'}{2}(a(x)-b(x)) +
\frac{y}{12}(a^3(x)+3a(x)b^2(x)) \nonumber\\
- \frac w2 \int_0^x
dz\,(a(x)-a(z))^2 -w \overline{a}a(x) 
\label{saddle1ab}\\
\fl
0 = (\tau-w p_d) b(x) + \frac{\tau-\tau'}{2}(a(x)-b(x)) +
\frac{y}{12}(b^3(x)+3b(x)a^2(x)) \nonumber\\
- \frac w2 \int_0^x dz\,
(b(x)-b(z))^2 - w \overline{b}b(x)
\label{saddle2ab}\\
\fl
q = p_d +\frac{2y}{3}p_d^3 - \frac w2 \int_0^1 dz\,(a^2(z)-b^2(z))
\label{saddle3ab}
\end{eqnarray}
and
\begin{eqnarray}
\fl
\beta f_\epsilon(q) =
qp_d - \frac{p_d^2}{2} - \frac{yp_d^4}{6} - \frac{q^2}{2(1-2\tau')} +
\frac{\tau}{2} \int_0^1 dz\,(a^2(z)+b^2(z)) - 
\frac{\tau-\tau'}{4} \int_0^1 dz\,(a(z)-b(z))^2\nonumber\\
 - \frac w6 \int_0^1 dz\,z(a^3(z)+b^3(z)) - 
\frac w2 \int_0^1 dz\int_z^1 dz'\,(a^2(z)a(z')+b^2(z)b(z'))\nonumber\\
 + \frac{y}{48}\int_0^1 dz\,(a^4(z)+6a^2(z)b^2(z)+b^4(z)) +
\frac{wp_d}{2}\int_0^1 dz\,(a^2(z)-b^2(z))
\label{freeab}
\end{eqnarray}
These saddle point equations can easily be solved for $\tau=\tau'=0$, and the
solution is 
\begin{eqnarray}
a(x) &= \left\{
\begin{array}{l@{\hspace{1cm}}l}
\frac{2wx}{y} & x<x'_2 \\
\frac{2wx'_2}{y} & x\ge x'_2
\end{array}
\right. \\
b(x) &= 0 \\
q &= p_d + \frac{2y}{3}p_d^3 - \frac{2w^3}{y^2}({x'}_2^2-\frac 23 {x'}_2^3) .
\label{pdat}
\end{eqnarray}
The breakpoint of $a(x)$ is $x'_2 = 1-\sqrt{1-\frac{yp_d}{w}}$. The value of
$p_d$ has to be obtained by solving Eq.~(\ref{pdat}) for $p_d$.

When we want to find the free energy perturbatively in the limit $0<|\tau'|\ll
q\ll 1$, we notice that $\Delta a(x)$ and $\Delta b(x)$ (the correction terms for
nonzero $\tau'$) are both of order $\tau'$. Since these
are the corrections to the saddle point solution found at $\tau'=0$, they only
contribute to the free energy to order $\tau'^2$ and can therefore be neglected
as long as we are only interested in the free energy to order $\tau'$. Plugging
the functions $a(x)$ and $b(x)$ just found into Eq.~(\ref{freeab}) yields
\begin{equation}
\beta f_\epsilon(q) = \frac w6 q^3 - \frac 34 \tau' q^2 + \mathcal O(\tau'^2)
+ \mathcal O(q^4)= \frac w6 q^3 + \frac{3\epsilon^2}{16}q^2 +\mathcal
O(\epsilon^4,q^4) \qquad(\epsilon^2\ll
q \ll 1).
\label{freeat2}
\end{equation}

We notice that for $\epsilon\ll N^{-1/6}$ the exponent $N\beta
f_\epsilon(q)$ of the probability distribution $P_\epsilon(q)\sim e^{-N\beta
f_\epsilon(q)}$ is dominated by $Nwq^3/6$ from Eq.~(\ref{freeat2}), while for
$N^{-1/6}\ll\epsilon\ll 1$ it is dominated by $N\frac{q^2}{2}h(\epsilon)$ from
Eq.~(\ref{freeat1}). Therefore the probability distribution is
\begin{equation}
P_\epsilon(q) \propto \left\{
\begin{array}{l@{\hspace{1cm}}l}
e^{-Nwq^3/6} & \epsilon\ll N^{-1/6} \\
e^{-Nq^2 h(\epsilon)/2} & N^{-1/6}\ll\epsilon
\end{array}
\right.
\end{equation}

\subsection{Below the critical temperature}

\subsubsection{Some exact solutions of the saddle point equations}

Eqs.~(\ref{saddle1})--(\ref{saddle3}) can not be solved for general $q$ and
$\epsilon$ (or rather $\tau'$ which contains the only reference to $\epsilon$).
They can however be solved exactly for $\epsilon=0$ and for $q=0$ and
perturbatively in various limits.

The solution for $q=0$ is simply
\begin{eqnarray}
p_1(x) &= 0 \\
q_1(x) &= \left\{
\begin{array}{l@{\hspace{1cm}}l}
\frac{wx}{2y} & x<x_2 \\
\frac{wx_2}{2y} & x\ge x_2
\end{array}
\right. \\
p_d&=0 
\label{solparisi}
\end{eqnarray}
where $x_2=1-\sqrt{1-\frac{4y\tau}{w^2}}$. That this is a solution can easily be
checked since Eq.~(\ref{saddle3}) is clearly satisfied, Eq.~(\ref{saddle2}) is
also satisfied by $p_d=0$ and $p(x)=0$, irrespective of $q(x)$, and
Eq.~(\ref{saddle1}) in this case reduces to the normal Parisi equation the
solution of which is well known and given in Eq.~(\ref{solparisi}).

The solution for $\tau=\tau'$ and $q<q_{\mathrm{EA}}$, where $q_{\mathrm{EA}}$
is the Edwards-Anderson order parameter, has been found by Rizzo
\cite{Rizzo:2001} and is given by
\begin{eqnarray}
p_2(x) &= \left\{
\begin{array}{l@{\hspace{1cm}}l}
\frac{wx}{y} & x<x_1 \\
\frac{wx_1}{y} & x\ge x_1
\end{array}
\right.
\label{taueqtaup2}\\
q_2(x) &= \left\{
\begin{array}{l@{\hspace{1cm}}l}
\frac{wx}{y} & x<x_1 \\
\frac{wx_1}{y} & x_1\le x < 2x_1 \\
\frac{wx}{2y} & 2x_1 \le x < x_2 \\
\frac{wx_2}{2y} & x_2 \le x \le 1
\end{array}
\right. \\
p_d &= \frac{q}{1-2\tau} \label{taueqtaup1}
\label{taueqtaup3}
\end{eqnarray}
with $x_1 = \frac{yp_d}{w}$. This solution however only exists if $2x_1<x_2$,
which is the case when $q<q_{\mathrm{EA}}$. In the limit $q\to 0$ it goes to the solution
found above for $q=0$. As shown by Rizzo, the free energy $\beta f_0$ of this 
solution (for any $q$ where this solution exists) is precisely equal to that 
of two unconstrained systems; we
may therefore use $\beta f_0$ as reference energy.

For $\tau=\tau'$ and $q>q_{\mathrm{EA}}$, another solution takes over
\cite{Billoire:2003}. It is given by
\begin{eqnarray}
p_3(x) &= q_3(x) = \left\{
\begin{array}{l@{\hspace{1cm}}l}
\frac{wx}{y} & x<x_3 \\
\frac{wx_3}{y} & x\ge x_3
\end{array}
\right. \\
q &= p_d + \frac{2y}{3}p_d^3 - \frac{2w^3}{y^2}(x_3^2-\frac 23 x_3^3) .
\label{pdbelow}
\end{eqnarray}
The breakpoint is $x_3 = 1-\sqrt{1-\frac{y\tau+wyp_d}{w^2}}$, and $p_d$ can be
obtained by solving Eq.~(\ref{pdbelow}) for $p_d$.
Since the details can be found in \cite{Billoire:2003}, we merely quote here
that the excess free energy of this solution is
\begin{equation}
\beta f-\beta f_0 = c_0(q-q_{\mathrm{EA}})^3
\label{freebeyondqEA}
\end{equation}
with some positive constant $c_0$.

\subsubsection{Perturbative solution in the limit $\epsilon\to\infty$}
The first limiting case to consider is $\epsilon\to\infty$ or, equivalently,
$\tau'\to-\infty$. In this case, dividing Eq.~(\ref{saddle2}) by $\tau'$ yields
to leading order
\begin{equation}
p(x) = \frac{w p_d q(x)}{-\tau'} + \mathcal O(1/\tau'^2).
\end{equation}
Inserting this into Eq.~(\ref{saddle1})  gives, again to leading order,
\begin{equation}
0 = (\tau+\frac{w^2p_d^2}{-\tau'})q(x) - \frac y3 q^3(x) - \frac w2\int_0^1 dz\,(q(x)-q(z))^2 - w\overline q q(x),
\end{equation}
i.e.\ precisely the Parisi equation but for a slightly \textit{lower} temperature $\tau+\frac{w^2p_d^2}{-\tau'}$.

Let's look at the free energy of this solution in the limit $\tau'=-\infty$. From
Eq.~(\ref{saddle3}) we get $q p_d=p_d^2+2yp_d^4/3$ and the free energy
difference $\beta f_\infty(q)-\beta f_0$ is
\begin{equation}
\beta f_\infty(q)-\beta f_0 = \frac{p_d^2}{2} + \frac{yp_d^4}{2} =
\frac{q^2}{2} + \mathcal O(q^4).
\end{equation}
This means that the probability density $P_\infty(q)$ is proportional to
$\exp(-N(q^2/2+\mathcal O(q^4)))$,
from which $[q^2]_0$ can be evaluated and is given by
$[q^2]_0 = \frac 1N$.
This is a useful result to check whether the method works since it is easy to
convice oneself that this is indeed true for two replicas with completely
uncorrelated couplings:
\begin{equation}
E\, \langle q^2\rangle = E\left\langle\left(\frac 1N \sum_{i}s_i t_i
 \right)^2\right\rangle 
 = E \left\langle\frac{2}{N^2} \left(\sum_{i<j}s_is_j t_it_j + \frac N2 \right)
 \right\rangle = \frac 1N,
\end{equation}
in agreement with the replica result.

\subsubsection{Perturbative solution for $\epsilon^2\ll q \ll 1$}

We now turn to the opposite limit, $\epsilon\to 0$ or $\tau'\to\tau$. We first
consider the case $\epsilon^2\ll q\ll 1$. In order to solve the saddle
point equations in this limit, it is again convenient to consider the version in terms
of $a(x)$ and $b(x)$, Eqs.~(\ref{saddle1ab})--(\ref{saddle3ab}). The solution for
$\tau'=\tau$, i.e.\ $\Delta\tau=\epsilon=0$, and arbitrary $q$ (small enough
such that the solution exists), taken from Eqs.~(\ref{taueqtaup1})--(\ref{taueqtaup3}), is
\begin{eqnarray}
p_d &= \frac{q}{1-2\tau} \\
a(x)&=\left\{
\begin{array}{l@{\hspace{1cm}}l}
\frac{2wx}{y}&x<x_1 \\
\frac{2wx_1}{y}&x_1 \le x < 2x_1 \\
\frac{w(x+2x_1)}{2y} & 2x_1\le x < x_2 \\
\frac{w(x_2+2x_1)}{2y} & x_2 \le x \le 1
\end{array}\right. \\
b(x) &= \left\{
\begin{array}{l@{\hspace{1cm}}l}
0 & x<2x_1 \\
\frac{w(x-2x_1)}{2y} & 2x_1 \le x < x_2 \\
\frac{w(x_2-2x_1)}{2y} & x_2 \le x \le 1
\end{array}
\right. .
\end{eqnarray}
Surprisingly, a perturbative expansion of the saddle point equations for small
$\Delta \tau$ shows that the leading order correction to this solution 
is of order $\sqrt{\Delta\tau}$. It is straightforward to show that it is given
by
\begin{equation}
\Delta b(x) = \left\{
\begin{array}{l@{\hspace{1cm}}l}
0 & x<x_1 \\
\sqrt{\frac{2\Delta\tau}{3y}} & x_1 < x < 2x_1 \\
0 & 2x_1 < x \le 1
\end{array}
\right. .
\end{equation}
Inserting this into the free energy difference leads to
\begin{equation}
\beta f_\epsilon(q)-\beta f_0 = \frac{yp_d^3}{3w}\Delta\tau =
\frac{y\epsilon^2 q^3}{12 w \beta^2(1-2\tau)^3} =: c_1\epsilon^2 q^3 \qquad
(\epsilon^2\ll q \ll 1),
\label{epsllq}
\end{equation}
defining the constant $c_1$.

\subsubsection{Perturbative solution for $q\ll\epsilon^2\ll 1$}

In this limit we can use the solution for
$q=0$ from Eq.~(\ref{solparisi}) as reference solution and construct a
perturbative solution for small $q$ (or $p_d$) from it, i.e.\ we set
$p(x)=p_1(x)+\Delta p(x)=\Delta p(x)$ and $q(x)=q_1(x)+\Delta q(x)$ and write
down the lowest order in $p_d$ from Eq.~(\ref{saddle2}):
\begin{equation}
\fl
0 = \tau' \Delta p(x) - w\int_0^x dz\,(\Delta p(x) - \Delta p(z))(q_1(x)-q_1(z)) - 
w(\overline{\Delta p}-p_d)q_1(x)-w\overline{q_1}\Delta p(x).
\label{saddle2b}
\end{equation}
As we can see, $\Delta q(x)$ does not appear in this equation at all, and since
$p(x)$ only appears quadratically or in combination with $p_d$ in
Eq.~(\ref{saddle1}), we conclude that $\Delta q(x)$ must be in fact of higher
order ($\mathcal O(p_d^2)$) such that we may set it equal to zero.

Eq.~(\ref{saddle2b}) can be solved in the following way. First note that
$\overline{q_1}=\tau/w$ such that with $\Delta \tau := \tau-\tau'$
\begin{equation}
\fl
	0 = -\Delta\tau \Delta p(x) - w\int_0^x dz\,(\Delta p(x) - \Delta p(z))(q_1(x)-q_1(z)) - w(\overline{\Delta p}-p_d)q_1(x).
	\label{saddle2c}
\end{equation}
Differentiating once with respect to $x$ yields 
\begin{eqnarray}
\fl
0 = -\Delta\tau \Delta p'(x) - wq'_1(x)\int_0^x dz\,(\Delta p(x) - \Delta
p(z)) - w\Delta p'(x)\int_0^x dz\,(q_1(x)-q_1(z)) \nonumber\\
 - w(\overline{\Delta p}-p_d)q'_1(x).
\label{saddle2d}
\end{eqnarray}
We conclude that $\Delta p'(x)=0$ whenever $q'_1(x)=0$. When $q'_1(x)\ne 0$,
differentiating Eq.~(\ref{saddle2d}) once more gives (noting that $q''_1(x)=0$)
\begin{eqnarray}
\fl
	0 &= -\Delta\tau \Delta p''(x) - 2wxq'_1(x)\Delta p'(x) - w\Delta
	p''(x)\int_0^x dz\,(q_1(x)-q_1(z)).
\\
\fl
 &= -\left(\Delta\tau+\frac{w^2x^2}{4y}\right)\Delta p''(x) -
 \frac{w^2x}{y}\Delta p'(x).
	\label{saddle2e}
\end{eqnarray}
This differential equation can easily be solved and one gets
\begin{equation}
\Delta p'(x) = C'\left(\Delta\tau + \frac{w^2x^2}{4y}\right)^{-2}
\end{equation}
with an as yet undetermined constant $C'$. Integrating once with
respect to $x$ yields
\begin{equation}
\Delta p(x) =
\frac{C'\sqrt{y}}{w\Delta\tau^{3/2}}f\left(\frac{wx}{2\sqrt{y\Delta\tau}}\right)
=: C f\left(\frac{wx}{2\sqrt{y\Delta\tau}}\right),
\end{equation}
where
\begin{equation} 
f(z) = \frac{z}{1+z^2}+\arctan z.
\label{deltap}
\end{equation}
This is true for $x<x_2$ while for $x>x_2$, $\Delta p(x)$ stays constant. We
can now calculate $\overline{\Delta p}$,
\begin{eqnarray}
\overline{\Delta p} &= C\int_0^{x_2}
dx\, f\left(\frac{wx}{2\sqrt{y\Delta\tau}}\right)
+C(1-x_2) f\left(\frac{wx_2}{2\sqrt{y\Delta\tau}}\right) \\
&=
C\arctan(wx_2/2\sqrt{y\Delta\tau}) +
C(1-x_2)\frac{wx_2/2\sqrt{y\Delta\tau}}{1+w^2x_2^2/4y\Delta\tau}.
\end{eqnarray}

The constant $C$ can be determined from Eq.~(\ref{saddle2d}) by taking the limit
$x\to 0$. In this limit the equation reads
\begin{equation}
\Delta\tau\Delta p'(0) = w(p_d-\overline{\Delta p})q'_1(0)
\end{equation}
such that
\begin{equation}
\frac{Cw\sqrt{\Delta\tau}}{\sqrt{y}} = \frac{w^2}{2y}p_d-\frac{w^2}{2y}\overline{\Delta
p} 
\end{equation}
which yields upon expanding in powers of
$\sqrt{\Delta\tau}$,
\begin{equation}
C = \frac{2}{\pi}p_d + \frac{4}{\pi^2}p_d\left(\frac 23-x_2\right)
\left(
\frac{2\sqrt{y\Delta\tau}}{wx_2}\right)^3+\mathcal O(\Delta\tau^{5/2}).
\end{equation}
With this value of $C$, $p(x)=\Delta p(x)$ together with $q(x)=q_1(x)$ is the
correct solution of Eqs.~(\ref{saddle1}) and (\ref{saddle2}) for small
$\Delta\tau$ up to order $p_d$.

What of the free energy of this solution? In order to calculate this, we
eliminate $q$ from Eq.~(\ref{free}) using Eq.~(\ref{saddle3}) and subtract
$\beta f_0$ to get
\begin{eqnarray}
\fl
	\beta f_\epsilon(q)-\beta f_0 = \frac{p_d^2}{2} + \frac{yp_d^4}{2} 
	-\frac{1}{2(1-2\tau')}\left(p_d + \frac{2yp_d^3}{3} -2w\int_0^1 dz\,p(z)q_1(z) \right)^2\nonumber\\ 
	 - \Delta\tau\int_0^1 dz\,p^2(z) + \frac y6 \int_0^1 dz\,p^4(z) - w\int_0^1 dz\,p^2(z)\int_0^z dz'(q_1(z)-q_1(z'))
	\nonumber\\
	-2w\int_0^1 dz\,p(z)q_1(z)\int_z^1 dz'\,p(z').
	\label{free2}
\end{eqnarray}
We can calculate the integrals appearing in this equation and obtain
\begin{eqnarray}
\int_0^1 dz\,p(z)q_1(z) &=& \frac{\tau p_d}{w} -\frac{\Delta\tau p_d}{w} +
\frac{4\sqrt{y}\Delta\tau^{3/2}p_d}{\pi w^2} +
\mathcal O(\Delta\tau^{5/2}) 
\\
\int_0^1 dz\, p^2(z) &=& p_d^2 - \frac{6\sqrt{y}}{\pi w}p_d^2\sqrt{\Delta\tau}+
\mathcal O(\Delta\tau^{3/2})
\end{eqnarray}
and
\begin{eqnarray}
\fl
\int_0^1 dz\, p(z)\left(p(z)\int_0^z dz'\,(q_1(z)-q_1(z')) +
2q_1(z)\int_z^1 dz'\,p(z')\right)  \nonumber\\
= \frac{\tau}{w}p_d^2 - 2\frac{\Delta\tau}{w}p_d^2 +
\frac{10\sqrt{y}\Delta\tau^{3/2}}{\pi w^2}p_d^2 + \mathcal O(\Delta \tau^{5/2}).
\end{eqnarray}
Assembling these results, the terms of order $\Delta\tau$ cancel and the free
energy difference is
\begin{eqnarray}
\fl
\beta f_\epsilon(q)-\beta f_0 = \frac{4\sqrt{y}}{\pi w}p_d^2\Delta\tau^{3/2} +
\mathcal O(\Delta\tau^3) = c_2 \epsilon^3 q^2 + \mathcal
O(\epsilon^5) \qquad\mathrm{with}\nonumber\\
 c_2 =\frac{4\sqrt{y}}{\pi
w}\frac{1}{(1-2\tau)^2}\frac{1}{8\beta^3}.
\label{qlleps}
\end{eqnarray}

\subsection{Perturbative solution for $q\ll \min(1,\epsilon^2)$}

In the preceding subsection we have calculated the free energy for
$q\ll\epsilon\ll 1$. The condition $\epsilon\ll 1$ is an unnecessary restriction,
however, and one can in principle carry out the same calculation as above without
expanding for small $\epsilon$, as long as $q\ll\min(1,\epsilon^2)$. The result
is
\begin{equation}
\beta f_\epsilon(q)-\beta f_0 = f(\epsilon) q^2 + \mathcal O(q^3),
\label{qll1eps}
\end{equation} 
where the function $f(\epsilon)$ is monotonic and has the properties
$f(\epsilon)=c_2 \epsilon^3 +\mathcal O(\epsilon^5)$ for $\epsilon\to 0$ and
$f(\epsilon)\to \frac 12$ for $\epsilon\to\infty$. The latter follows from the
solution for $\tau'\to-\infty$ we found previously. As we won't need any more
detailed information about $f(\epsilon)$, we will not carry out this tedious
calculation in detail here.

\subsection{Probability distribution}

As before, we want to calculate the probability distribution $P_\epsilon^0(q)\sim
e^{-N\beta (f_\epsilon(q)-f_0)}$. We observe that the admissible range of
$\epsilon$ (the interval from $0$ to $\infty$) divides into $4$ parts. For
$\epsilon\ll N^{-1/2}$, both Eq.~(\ref{epsllq}) and (\ref{qlleps}) produce a
negligible exponent $N\beta(f_\epsilon(q)-f_0)$ for all $q\in
[0,q_{\mathrm{EA}}]$. The probability distribution $P_\epsilon(q)$ is thus a
constant in that interval, with an exponentially suppressed tail for
$q>q_{\mathrm{EA}}$ from Eq.~(\ref{freebeyondqEA}). In the range
$N^{-1/2}\ll\epsilon\ll N^{-1/5}$ the exponent $N\beta(f_\epsilon(q)-f_0)$ is
negligible for $q \ll \epsilon^2$ when Eq.~(\ref{qlleps}) prevails. It only
becomes noticable when $q\ge N^{-2/5}\gg \epsilon^2$ such that we can approximate
it by Eq.~(\ref{epsllq}). For $N^{-1/5}\ll \epsilon \le \epsilon_0$ with some
arbitrary small and fixed $\epsilon_0$ independent of $N$, the exponent is
dominated by Eq.~(\ref{qlleps}). Finally, for $\epsilon_0<\epsilon$ we can use
Eq.~(\ref{qll1eps}).

Combined, we can write
\begin{equation}
P_\epsilon(q) \propto \left\{
\begin{array}{l@{\hspace{1cm}}l}
\hat\theta(q-q_{\mathrm{EA}}) & \epsilon\ll N^{-1/2} \\
e^{-Nc_1 \epsilon^2 q^3} & N^{-1/2}\ll\epsilon\ll N^{-1/5} \\
e^{-Nc_2 \epsilon^3 q^2} & N^{-1/5}\ll\epsilon\le\epsilon_0 \\
e^{-Nf(\epsilon) q^2} & \epsilon_0<\epsilon
\end{array}
\right.,
\label{probbelow}
\end{equation}
where the function $\hat\theta(x) = 1$ for $x<0$ and $\hat\theta(x)=e^{-Nc_0x^3}$
for $x>0$. In principle, the latter two regimes could be combined to
$P_\epsilon(q)\propto e^{-Nf(\epsilon)q^2}$ for $N^{-1/5}\ll\epsilon$ but
splitting them makes the dependence on $\epsilon$ more explicit.

A note is in order about the probability distribution at $\epsilon=0$. Clearly,
the result in Eq.~(\ref{probbelow}) does not coincide with the known distribution
from the Parisi solution \cite{Parisi:1979,Parisi:1980a,Parisi:1980b}, in
particular the $\delta$ peak at $q_{\mathrm{EA}}$ is missing and in the range
$0\le q<q_{\mathrm{EA}}$ the probability distribution is flat. The latter is not
only an artifact of using the truncated model (for which the distribution would
indeed be flat) but was already derived in \cite{Rizzo:2001,Billoire:2003} for
the full model. Both the flatness and the missing $\delta$ peak originate from
neglecting the finite size corrections in Eq.~(\ref{probability}). We conclude
that these corrections are of order $1$ for $q<q_{\mathrm{EA}}$ (but do not
change the probability distribution \textit{qualitatively}, i.e.\ do not
introduce gaps or zeros in the distribution) and conspire to form the $\delta$
peak for large $N$ at $q=q_{\mathrm{EA}}$. When $\epsilon>N^{-1/2}$ we do not
expect any $\delta$ peaks since the equilibrium states in the two replicas start
to differ substantially, and in the light of what we saw at $\epsilon=0$ we
expect the finite size corrections to be similarly good-natured. Thus
the results we have presented here for the probability distribution should be
correct up to prefactors which might vary for the full model.

%\section{Comparison with computer simulations}
%\label{secnumerical}

\section{Conclusion}
\label{secconclusion}

We have shown that bond chaos exists in the Sherrington-Kirkpatrick model by
calculating the probability distribution of the overlap $q$ between two copies of
a system, one of which has randomly perturbed bonds with respect to the other.
The finite size scaling of this distribution has been calculated above, at and
below the critical temperature. In the low temperature phase four different
regimes have been identified. For bond distances $\epsilon\ll N^{-1/2}$ the
distribution has a variance of order $1$, i.e.\ the equilibrium states in the two
copies are still very similar. For $N^{-1/2}\ll\epsilon\ll N^{-1/5}$, the
distribution is proportional to $e^{-Nc_1\epsilon^2q^3}$, i.e.\ already very
narrow, the width scaling as $\xi_1^{-2/3}$ with the scaling variable
$\xi_1=\sqrt{N}\epsilon$. For $N^{-1/5}\ll\epsilon\ll 1$, the scaling variable is
$\xi_2=N^{1/3}\epsilon$ and the distribution becomes Gaussian with width
proportional to $\xi_2^{-3/2}$. For all other values of $\epsilon$, finally, the
distribution remains Gaussian, and its width goes as $N^{-1/2}$.

\ack
I would like to thank M.A.\ Moore, M.\ Goethe and A.\ Braun for discussions and
T.\ Rizzo for an exchange of emails.

\section*{References}

\bibliographystyle{unsrt}
\bibliography{LiteraturDB,cond-mat}

\end{document}